\pdfoutput=1
\documentclass[12pt,a4paper]{article}
\usepackage{ifthen}
\newboolean{pdflatex}
\setboolean{pdflatex}{true}

\newboolean{articletitles}
\setboolean{articletitles}{true}

\newboolean{uprightparticles}
\setboolean{uprightparticles}{false}

\newboolean{inbibliography}
\setboolean{inbibliography}{false}

\usepackage[top=1in, bottom=1.25in, left=1in, right=1in]{geometry}

\usepackage{microtype}
\usepackage{lineno}
\usepackage{xspace}
\usepackage{caption}

\usepackage{graphicx}
\usepackage{color}
\usepackage{colortbl}

\usepackage{amsmath}
\usepackage{amssymb}
\usepackage{amsfonts}
\usepackage{upgreek}

\newcommand*\patchAmsMathEnvironmentForLineno[1]{
\expandafter\let\csname old#1\expandafter\endcsname\csname #1\endcsname
\expandafter\let\csname oldend#1\expandafter\endcsname\csname
end#1\endcsname
 \renewenvironment{#1}
   {\linenomath\csname old#1\endcsname}
   {\csname oldend#1\endcsname\endlinenomath}
}
\newcommand*\patchBothAmsMathEnvironmentsForLineno[1]{
  \patchAmsMathEnvironmentForLineno{#1}
  \patchAmsMathEnvironmentForLineno{#1*}
}
\AtBeginDocument{
\patchBothAmsMathEnvironmentsForLineno{equation}
\patchBothAmsMathEnvironmentsForLineno{align}
\patchBothAmsMathEnvironmentsForLineno{flalign}
\patchBothAmsMathEnvironmentsForLineno{alignat}
\patchBothAmsMathEnvironmentsForLineno{gather}
\patchBothAmsMathEnvironmentsForLineno{multline}
\patchBothAmsMathEnvironmentsForLineno{eqnarray}
}

\usepackage{hyperref}
\usepackage[all]{hypcap}

\usepackage{xspace}
\usepackage{upgreek}

\def\lhcb {\mbox{LHCb}\xspace}

\def\cdf    {\mbox{CDF}\xspace}
\def\dzero  {\mbox{D0}\xspace}

\def\MagUp {\mbox{\em Mag\kern -0.05em Up}\xspace}

\ifthenelse{\boolean{uprightparticles}}
{

 \def\Ppi         {\ensuremath{\uppi}\xspace}

 \def\Ppsi        {\ensuremath{\uppsi}\xspace}

 \def\PDelta      {\ensuremath{\Delta}\xspace}
 \def\PXi      {\ensuremath{\Xi}\xspace}
 \def\PLambda      {\ensuremath{\Lambda}\xspace}
 \def\PSigma      {\ensuremath{\Sigma}\xspace}
 \def\POmega      {\ensuremath{\Omega}\xspace}
 \def\PUpsilon      {\ensuremath{\Upsilon}\xspace}

 \def\PB      {\ensuremath{\mathrm{B}}\xspace}
 
 \def\PD      {\ensuremath{\mathrm{D}}\xspace}

 \def\PJ      {\ensuremath{\mathrm{J}}\xspace}
 \def\PK      {\ensuremath{\mathrm{K}}\xspace}

 \def\Pb      {\ensuremath{\mathrm{b}}\xspace}
 \def\Pc      {\ensuremath{\mathrm{c}}\xspace}

 \def\Pi      {\ensuremath{\mathrm{i}}\xspace}

 \def\Pp      {\ensuremath{\mathrm{p}}\xspace}

 \def\Ps      {\ensuremath{\mathrm{s}}\xspace}
 \def\Pt      {\ensuremath{\mathrm{t}}\xspace}

}
{

 \def\Ppi         {\ensuremath{\pi}\xspace}

 \def\Ppsi        {\ensuremath{\psi}\xspace}
 
 \mathchardef\PDelta="7101
 \mathchardef\PXi="7104
 \mathchardef\PLambda="7103
 \mathchardef\PSigma="7106
 \mathchardef\POmega="710A
 \mathchardef\PUpsilon="7107
 
 \def\PB      {\ensuremath{B}\xspace}
 
 \def\PD      {\ensuremath{D}\xspace}

 \def\PJ      {\ensuremath{J}\xspace}
 \def\PK      {\ensuremath{K}\xspace}

 \def\Pb      {\ensuremath{b}\xspace}
 \def\Pc      {\ensuremath{c}\xspace}

 \def\Pi      {\ensuremath{i}\xspace}

 \def\Pp      {\ensuremath{p}\xspace}

 \def\Ps      {\ensuremath{s}\xspace}
 \def\Pt      {\ensuremath{t}\xspace}

}

\makeatletter
\ifcase \@ptsize \relax
  \newcommand{\miniscule}{\@setfontsize\miniscule{4}{5}}
\or
  \newcommand{\miniscule}{\@setfontsize\miniscule{5}{6}}
\or
  \newcommand{\miniscule}{\@setfontsize\miniscule{5}{6}}
\fi
\makeatother

\DeclareRobustCommand{\optbar}[1]{\shortstack{{\miniscule (\rule[.5ex]{1.25em}{.18mm})}
  \\ [-.7ex] $#1$}}

\def\squark    {{\ensuremath{\Ps}}\xspace}

\def\cquark    {{\ensuremath{\Pc}}\xspace}
\def\cquarkbar {{\ensuremath{\overline \cquark}}\xspace}

\def\bquark    {{\ensuremath{\Pb}}\xspace}
\def\bquarkbar {{\ensuremath{\overline \bquark}}\xspace}

\def\tquark    {{\ensuremath{\Pt}}\xspace}

\def\pion   {{\ensuremath{\Ppi}}\xspace}

\def\pip    {{\ensuremath{\pion^+}}\xspace}
\def\pim    {{\ensuremath{\pion^-}}\xspace}

\def\kaon    {{\ensuremath{\PK}}\xspace}
  \def\Kbar    {{\kern 0.2em\overline{\kern -0.2em \PK}{}}\xspace}

\def\KorKbar    {\kern 0.18em\optbar{\kern -0.18em K}{}\xspace}

\def\Kp      {{\ensuremath{\kaon^+}}\xspace}
\def\Km      {{\ensuremath{\kaon^-}}\xspace}

  \def\Dbar    {{\kern 0.2em\overline{\kern -0.2em \PD}{}}\xspace}

\def\DorDbar    {\kern 0.18em\optbar{\kern -0.18em D}{}\xspace}

\def\B       {{\ensuremath{\PB}}\xspace}
\def\Bbar    {{\ensuremath{\kern 0.18em\overline{\kern -0.18em \PB}{}}}\xspace}

\def\BorBbar    {\kern 0.18em\optbar{\kern -0.18em B}{}\xspace}

\def\Bu      {{\ensuremath{\B^+}}\xspace}

\def\Bd      {{\ensuremath{\B^0}}\xspace}
\def\Bs      {{\ensuremath{\B^0_\squark}}\xspace}
\def\Bsb     {{\ensuremath{\Bbar{}^0_\squark}}\xspace}

\def\jpsi     {{\ensuremath{{\PJ\mskip -3mu/\mskip -2mu\Ppsi\mskip 2mu}}}\xspace}
\def\psitwos  {{\ensuremath{\Ppsi{(2S)}}}\xspace}

  \def\Y#1S{\ensuremath{\PUpsilon{(#1S)}}\xspace}

\def\proton      {{\ensuremath{\Pp}}\xspace}

\def\Lz          {{\ensuremath{\PLambda}}\xspace}
\def\Lbar        {{\ensuremath{\kern 0.1em\overline{\kern -0.1em\PLambda}}}\xspace}
\def\LorLbar    {\kern 0.18em\optbar{\kern -0.18em \PLambda}{}\xspace}

\def\Lb      {{\ensuremath{\Lz^0_\bquark}}\xspace}

\newcommand{\decay}[2]{\ensuremath{#1\!\to #2}\xspace}

\def\to                 {\ensuremath{\rightarrow}\xspace}

\def\CP                {{\ensuremath{C\!P}}\xspace}

\def\Vcs  {{\ensuremath{V_{\cquark\squark}}}\xspace}
\def\Vts  {{\ensuremath{V_{\tquark\squark}}}\xspace}

\def\Vcbs  {{\ensuremath{V_{\cquark\bquark}^\ast}}\xspace}
\def\Vtbs  {{\ensuremath{V_{\tquark\bquark}^\ast}}\xspace}

\newcommand{\dms}{{\ensuremath{\Delta m_{\squark}}}\xspace}

\newcommand{\DGs}{{\ensuremath{\Delta\Gamma_{\squark}}}\xspace}

\newcommand{\Gs}{{\ensuremath{\Gamma_{\squark}}}\xspace}

\newcommand{\GL}{{\ensuremath{\Gamma_{\mathrm{ L}}}}\xspace}
\newcommand{\GH}{{\ensuremath{\Gamma_{\mathrm{ H}}}}\xspace}

\newcommand{\phis}{{\ensuremath{\phi_{\squark}}}\xspace}
\newcommand{\betas}{{\ensuremath{\beta_{\squark}}}\xspace}

\newcommand{\mistag}{\ensuremath{\omega}\xspace}

\def\BsToJPsiPhi  {\decay{\Bs}{\jpsi\phi}}
\def\BsToPsiTwoSPhi  {\decay{\Bs}{\psi(2S)\phi}}

\def\AT#1     {\ensuremath{A_{\mathrm{T}}^{#1}}\xspace}

\def\C#1      {\ensuremath{\mathcal{C}_{#1}}\xspace}
\def\Cp#1     {\ensuremath{\mathcal{C}_{#1}^{'}}\xspace}
\def\Ceff#1   {\ensuremath{\mathcal{C}_{#1}^{\mathrm{(eff)}}}\xspace}
\def\Cpeff#1  {\ensuremath{\mathcal{C}_{#1}^{'\mathrm{(eff)}}}\xspace}
\def\Ope#1    {\ensuremath{\mathcal{O}_{#1}}\xspace}
\def\Opep#1   {\ensuremath{\mathcal{O}_{#1}^{'}}\xspace}

\newcommand{\ket}[1]{\ensuremath{|#1\rangle}}

\newcommand{\tev}{\ifthenelse{\boolean{inbibliography}}{\ensuremath{~T\kern -0.05em eV}\xspace}{\ensuremath{\mathrm{\,Te\kern -0.1em V}}}\xspace}
\newcommand{\gev}{\ensuremath{\mathrm{\,Ge\kern -0.1em V}}\xspace}
\newcommand{\mev}{\ensuremath{\mathrm{\,Me\kern -0.1em V}}\xspace}
\newcommand{\kev}{\ensuremath{\mathrm{\,ke\kern -0.1em V}}\xspace}
\newcommand{\ev}{\ensuremath{\mathrm{\,e\kern -0.1em V}}\xspace}
\newcommand{\gevc}{\ensuremath{{\mathrm{\,Ge\kern -0.1em V\!/}c}}\xspace}
\newcommand{\mevc}{\ensuremath{{\mathrm{\,Me\kern -0.1em V\!/}c}}\xspace}
\newcommand{\gevcc}{\ensuremath{{\mathrm{\,Ge\kern -0.1em V\!/}c^2}}\xspace}
\newcommand{\gevgevcccc}{\ensuremath{{\mathrm{\,Ge\kern -0.1em V^2\!/}c^4}}\xspace}
\newcommand{\mevcc}{\ensuremath{{\mathrm{\,Me\kern -0.1em V\!/}c^2}}\xspace}

\def\mum  {\ensuremath{{\,\upmu\mathrm{m}}}\xspace}

\def\invfb   {\ensuremath{\mbox{\,fb}^{-1}}\xspace}

\def\ps   {\ensuremath{{\mathrm{ \,ps}}}\xspace}
\def\fs   {\ensuremath{\mathrm{ \,fs}}\xspace}

\def\invps{\ensuremath{{\mathrm{ \,ps^{-1}}}}\xspace}

\def\deriv {\ensuremath{\mathrm{d}}}

\def\gsim{{~\raise.15em\hbox{$>$}\kern-.85em
          \lower.35em\hbox{$\sim$}~}\xspace}
\def\lsim{{~\raise.15em\hbox{$<$}\kern-.85em
          \lower.35em\hbox{$\sim$}~}\xspace}

\def\ptot       {\mbox{$p$}\xspace}
\def\pt         {\mbox{$p_{\mathrm{ T}}$}\xspace}

\def\rad{\ensuremath{\mathrm{ \,rad}}\xspace}

\def\evtgen     {\mbox{\textsc{EvtGen}}\xspace}

\def\geant      {\mbox{\textsc{Geant4}}\xspace}

\def\photos     {\mbox{\textsc{Photos}}\xspace}

\def\pythia     {\mbox{\textsc{Pythia}}\xspace}

\def\tell1  {TELL1\xspace}
\def\ukl1   {UKL1\xspace}

\usepackage{cite}
\usepackage{mciteplus}
\usepackage{longtable}

\usepackage{cite}
\usepackage{mciteplus}
\usepackage{verbatim}
\begin{document}

\renewcommand{\thefootnote}{\fnsymbol{footnote}}
\setcounter{footnote}{1}

\begin{titlepage}
\pagenumbering{roman}

\vspace*{-1.5cm}
\centerline{\large EUROPEAN ORGANIZATION FOR NUCLEAR RESEARCH (CERN)}
\vspace*{1.5cm}
\noindent
\begin{tabular*}{\linewidth}{lc@{\extracolsep{\fill}}r@{\extracolsep{0pt}}}
\vspace*{-2.7cm}\mbox{\!\!\!\includegraphics[width=.14\textwidth]{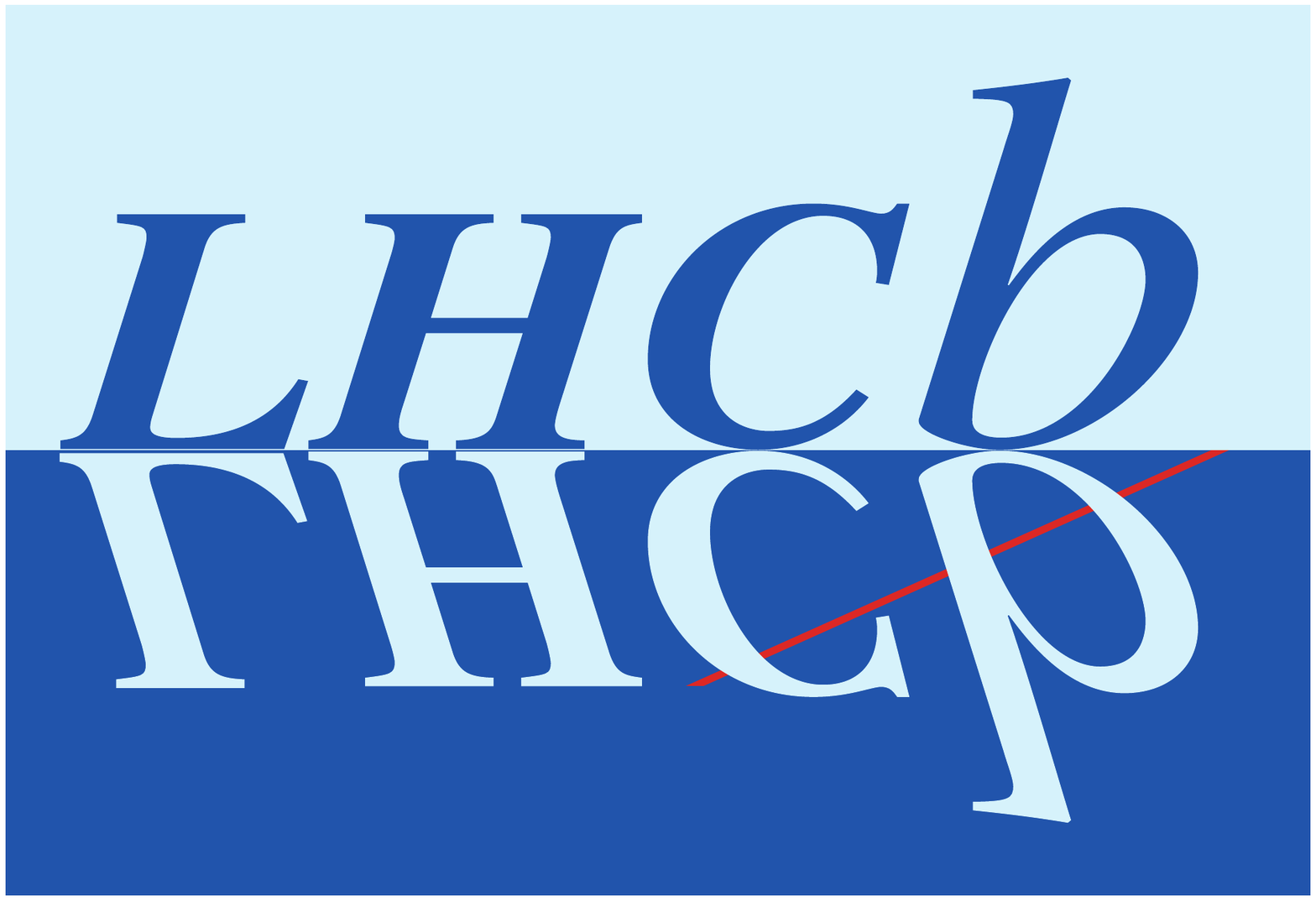}} & &
\\
 & & CERN-EP-2016-192 \\
 & & LHCb-PAPER-2016-027 \\
 & & September 15, 2016 \\
 & & \\
\end{tabular*}

\vspace*{2.0cm}

{\normalfont\bfseries\boldmath\huge
\begin{center}
  First study of the \CP-violating phase and decay-width difference in $B_s^0\to\psi(2S)\phi$ decays
\end{center}
}

\vspace*{1.5cm}

\begin{center}
The LHCb collaboration\footnote{Authors are listed at the end of this paper.}
\end{center}

\vspace{\fill}

\begin{abstract}
  \noindent
  A time-dependent angular analysis of $B_s^0\to\psi(2S)\phi$ decays
  is performed using data  recorded by the LHCb experiment.
  The data set corresponds to an integrated luminosity of 3.0\invfb collected during Run 1 of the LHC.
  The \CP-violating phase and decay-width difference of the \Bs system are
  measured to be $\phis = 0.23^{+0.29}_{-0.28} \pm 0.02$\rad and
  $\Delta\Gamma_s = 0.066^{+0.041}_{-0.044} \pm 0.007$\invps, respectively,
  where the first uncertainty is statistical and the second systematic.
  This is the first time that $\phi_s$ and $\Delta\Gamma_s$ have been
  measured in a decay containing the $\psi(2S)$ resonance.
\end{abstract}

\vspace*{2.0cm}

\begin{center}
  Published in Phys.~Lett.~B
\end{center}

\vspace{\fill}

{\footnotesize
\centerline{\copyright~CERN on behalf of the \lhcb collaboration, licence \href{http://creativecommons.org/licenses/by/4.0/}{CC-BY-4.0}.}}
\vspace*{2mm}

\end{titlepage}

\newpage
\setcounter{page}{2}
\mbox{~}

\cleardoublepage

\renewcommand{\thefootnote}{\arabic{footnote}}
\setcounter{footnote}{0}

\pagestyle{plain}
\setcounter{page}{1}
\pagenumbering{arabic}

\section{Introduction}
\label{sec:Introduction}

The interference between the amplitudes of decays of \Bs mesons
to $\cquark \cquarkbar X$ \CP eigenstates directly or via mixing, gives rise to a
\CP-violating phase, $\phis$.
In the Standard Model (SM),
ignoring subleading penguin contributions,
this phase is predicted to be $-2\betas$,
where $\betas = \arg[-(\Vts \Vtbs)/(\Vcs \Vcbs)]$ and $\ensuremath{V_{ij}}$ are elements of the CKM quark
flavour mixing matrix~\cite{Kobayashi:1973fv,*Cabibbo:1963yz}.

Measurements of $\phis$ using $\decay{\Bs}{\jpsi \Kp \Km}$ and $\decay{\Bs}{\jpsi \pip \pim}$
decays have been reported previously by the LHCb collaboration~\cite{LHCb-PAPER-2014-059}
based upon $3.0\invfb$ of integrated luminosity collected in $\proton \proton$ collisions at a
centre-of-mass energy of 7\tev in 2011 and  8\tev in 2012 at the LHC. Measurements of $\phis$ using \BsToJPsiPhi{} decays
have also been made by the \dzero~\cite{Abazov:2011ry}, \cdf~\cite{CDF:2011af}, CMS~\cite{Khachatryan:2015nza}
and ATLAS~\cite{Aad:2016tdj} collaborations.
The world-average value of these direct measurements is $\phis = -0.033\pm 0.033$\rad~\cite{HFAG}.
The global average from indirect measurements
gives $\phis = -0.0376^{+0.0007}_{-0.0008}$ rad~\cite{Charles:2015gya}.
Measurements of $\phis$ are interesting since new physics (NP) processes
could modify the phase if new particles were to contribute to the
box diagrams describing $\Bs$--$\Bsb$ mixing~\cite{Buras:2009if,Chiang:2009ev}.

In this analysis $\phis$ is measured using a flavour tagged, decay-time dependent angular
analysis of $\decay{\Bs}{\psi(2S) \phi}$ decays, with $\psi(2S)\to\mu^+\mu^-$ and $\phi\to\Kp\Km$.
In addition, measurements of the decay-width difference of the light (L) and heavy (H)
$\Bs$ mass eigenstates, $\DGs \equiv \GL - \GH$, the average $\Bs$ decay width, $\Gs \equiv (\GL + \GH)/2$,
and the polarisation amplitudes of the $\decay{\Bs}{\psi(2S) \phi}$ decay are reported.
This is the first time that a higher $c\overline{c}$ resonance is used to measure $\phis$.

This analysis follows very closely that of $\decay{\Bs}{\jpsi \Kp \Km}$ decays in
Refs.~\cite{LHCb-PAPER-2014-059,LHCb-PAPER-2013-002},
and only significant changes with respect to those analyses are described in this paper.
Section~\ref{sec:phenomenology} describes the phenomenology of the $\decay{\Bs}{\psi(2S) \phi}$ decay
and the physics observables.
Section~\ref{sec:Selection} describes the LHCb detector, data
and simulated samples that are used along with the optimisation of their selection.
Section~\ref{sec:Resolution} details the \Bs meson decay-time resolution,
decay-time efficiency and angular acceptance and  Section~\ref{sec:tagging} describes the flavour tagging algorithms.
Results and systematic uncertainties are given in Section~\ref{sec:Results} and
Section~\ref{sec:Systematics}, respectively. Conclusions are presented in Section~\ref{sec:conclusions}.

\section{Phenomenology}
\label{sec:phenomenology}
\def\assq{|A_{s}|^2}
\def\apar{A_{\|}}
\def\aperp{A_{\perp}}
\def\aparsq{|A_{\|}|^2}
\def\aperpsq{|A_{\perp}|^2}
\def\assq{|A_{s}|^2}
\def\azero{A_{0}(t)}
\def\azerosq{|A_{0}|^2}
\def\Rt{R_{\perp}}
\def\Rp{R_{0}}
\def\dgam{\Delta\Gamma_s}
\def\bh{B_H}
\def\bl{B_L}
\def\bs{B_s}
\def\bsbar{ \bar{B_s}}
\def\gaml{\Gamma_L}
\def\gamh{\Gamma_H}
\def\gambar{\bar\Gamma_s}
\def\delone{\delta_1}
\def\deltwo{\delta_2}
\def\dels{\delta_S}
\def\delone{\delta_1}
\def\deltwo{\delta_2}
\def\delpar{\delta_\parallel}
\def\delperp{\delta_\perp}
\def\delzero{\delta_0}
\def\dels{\delta_{\rm S}}
\def\delsperp{\dels - \delperp}

\def\maglambda{|\lambda|}

\begin{figure}[t]
\begin{center}
    \includegraphics[scale=1.1, clip=true, trim=30mm 215.5mm 0mm 32.5mm]{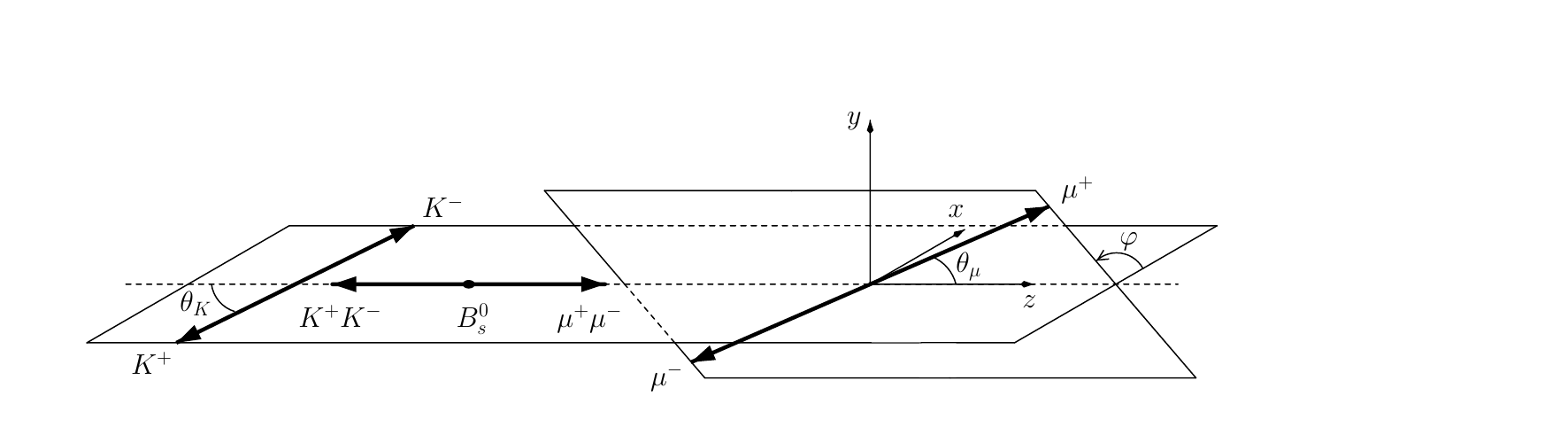}
\end{center}
  \caption{\small Definition of helicity angles.}
  \label{fig:helicity}
\end{figure}

The full formalism used for this analysis can be found in Ref.~\cite{LHCb-PAPER-2013-002}, where the
\jpsi is now replaced with the $\psi(2S)$ meson.
The differential cross-section as a function of the signal decay time, $t$, and three helicity
angles, $\Omega = (\cos\theta_\mu, \cos\theta_K, \varphi)$ (Fig.~\ref{fig:helicity}), is described by a
sum of ten terms, corresponding to the four polarisation amplitudes
(three corresponding to the  $\Kp\Km$ from the $\phi$ being in a $P$-wave configuration, and
one to allow for an additional non-resonant $\Kp\Km$ $S$-wave component) and their interference terms.
Each term is the product of a
time-dependent function and an angular function,
\begin{equation}
  X(t, \Omega) \equiv \frac{\deriv^{4} \Gamma(\BsToPsiTwoSPhi) }{\deriv t \;\deriv\Omega} \; \propto \;
  \sum^{10}_{k=1} \: h_k(t) \: f_k( \Omega) \,,
  \label{eq:Eqbsrate}
\end{equation}
where the definitions of $h_k(t)$ and $f_k( \Omega)$ are given in Ref.~\cite{LHCb-PAPER-2013-002}.
The $f_k( \Omega)$ functions depend only upon the final-state decay angles.
The $h_k(t)$ functions depend upon all physics parameters of interest, which are
$\Gs$,  $\DGs$, $\phis$, $\maglambda$, the mass difference of the \Bs eigenstates, \dms,
and the polarisation amplitudes $A_i = |A_i|e^{-i\delta_i}$,
where the indices $i\in\{0, \parallel, \perp, {S}\}$ refer to the different polarisation states of the
$\Kp\Km$ system.
The sum $\aparsq+\azerosq+\aperpsq$ equals unity and by convention $\delzero$ is zero.
The $S$-wave fraction is defined as
\mbox{$F_{S} \equiv |A_{S}|^2/(|A_{0}|^2 + |A_{\perp}|^2 + |A_{\parallel}|^2 + |A_{S}|^2)$}.
The parameter $\lambda$ describes \CP{} violation in the interference between
mixing and decay and is defined by $\lambda = \eta_i (q/p) (\bar{A}_i/A_i)$.
The complex parameters $p = \langle\Bs | B_{s,{\rm L}}\rangle$ and $q = \langle\Bsb | B_{s,{\rm L}}\rangle$
describe the relation between flavour and mass eigenstates,
where $B_{s,{\rm L}}$ is the light mass eigenstate and  $\eta_i$ is the \CP{} eigenvalue
of the polarisation state $i$.
The \CP-violating phase is defined by $\phis \equiv - \arg{(\eta_i \lambda)}$ and is assumed here
to be the same for all polarisation states.
In the absence of \CP{} violation in decay it follows that $\maglambda=1$.
In this paper \CP violation in \Bs-meson mixing is assumed to be negligible,
following measurements in Refs.~\cite{LHCb-PAPER-2013-033,LHCb-PAPER-2016-013}.

\section{Detector, data set and selection}
\label{sec:Selection}

The \lhcb detector~\cite{Alves:2008zz,LHCb-DP-2014-002} is a single-arm forward
spectrometer covering the \mbox{pseudorapidity} range $2<\eta <5$,
designed for the study of particles containing \bquark or \cquark
quarks. The detector includes a high-precision tracking system
consisting of a silicon-strip vertex detector surrounding the $pp$
interaction region, a large-area silicon-strip detector located
upstream of a dipole magnet with a bending power of about
$4{\mathrm{\,Tm}}$, and three stations of silicon-strip detectors and straw
drift tubes placed downstream of the magnet.
The tracking system provides a measurement of momentum, \ptot, of charged particles with
a relative uncertainty that varies from 0.5\% at low momentum to 1.0\% at 200\gevc.
The minimum distance of a track to a primary vertex (PV), the impact parameter, is measured with a resolution of $(15+29/\pt)\mum$,
where \pt is the component of the momentum transverse to the beam, in\,\gevc.
Different types of charged hadrons are distinguished using information
from two ring-imaging Cherenkov detectors.
Photons, electrons and hadrons are identified by a calorimeter system consisting of
scintillating-pad and preshower detectors, an electromagnetic
calorimeter and a hadronic calorimeter. Muons are identified by a
system composed of alternating layers of iron and multiwire
proportional chambers.

The online event selection is performed by a trigger~\cite{LHCb-DP-2012-004},
which consists of a
hardware stage, based on information from the calorimeter and the muon
system, followed by a software stage.
In this analysis, candidates are required to pass the
hardware trigger that selects muons and muon pairs based
on their transverse momentum. In the software stage, events
are triggered by a $\psi(2S) \to \mu^+\mu^-$ candidate, where the
$\psi(2S)$ is required to be consistent with coming from the decay
of a $b$ hadron, by using either impact parameter requirements on
the decay products or the detachment of the $\psi(2S)$ candidate from the PV.

In the simulation, $pp$ collisions are generated using
\pythia~\cite{Sjostrand:2006za,*Sjostrand:2007gs}
 with a specific \lhcb
configuration~\cite{LHCb-PROC-2010-056}.  Decays of hadronic particles
are described by \evtgen~\cite{Lange:2001uf}, in which final-state
radiation is generated using \photos~\cite{Golonka:2005pn}. The
interaction of the generated particles with the detector, and its response,
are implemented using the \geant
toolkit~\cite{Allison:2006ve, *Agostinelli:2002hh} as described in
Ref.~\cite{LHCb-PROC-2011-006}.

The $\Bs \to \psi(2S) \phi$ candidates are first selected with loose requirements to
ensure high efficiency and significant background rejection. The $\psi(2S)$ candidates are reconstructed
from pairs of oppositely-charged particles identified as muons, and the $\phi$
candidates are reconstructed from pairs of oppositely-charged particles identified as kaons.
The invariant mass of the
muon (kaon) pair must be within $60\mevcc$ ($12\mevcc$) of the known $\psi(2S)$ ($\phi$)
mass~\cite{PDG2014}.
Reconstructed kaon tracks that do not correspond to actual trajectories of charged particles
are suppressed by requiring a good track $\chi^2$ per degree of freedom. The $\pt$ of each
$\phi$ candidate is required to be larger than 1\gevc.

The $\psi(2S)$ and $\phi$ candidates
that are consistent with originating from a common vertex
are combined to create \Bs candidates.
Subsequently, a kinematic fit~\cite{Hulsbergen:2005pu} is applied to the \Bs candidates
in which the $\psi(2S)$ mass is constrained to the known value~\cite{PDG2014}
and the \Bs candidate is required to point back to the PV, to improve
the resolution on the invariant mass $m(\psi(2S)\Kp\Km)$.
Combinatorial background from particles produced at the PV is reduced by requiring that the
\Bs candidate decay time (computed from a vertex fit without the PV constraint) is larger than 0.3\ps.
Backgrounds from the misidentification of final-state particles from other decays such as $\Bd\to\psi(2S)K^+\pi^-$
and $\Lb\to\psi(2S)pK^-$ are negligible.

To further improve the signal-to-background ratio, a boosted decision tree
(BDT)~\cite{Breiman,AdaBoost} is applied.
The BDT is trained using simulated $\Bs\to\psi(2S)\phi$ events for the signal, while candidates from
data with $m(\psi(2S)\Kp\Km)$ larger than 5400\mevcc are used to model the background.
Twelve variables that have good discrimination power between signal and background are
used to define and train the BDT. These are: the \Bs candidate kinematic fit $\chi^2$;
the \pt of the \Bs and $\phi$ candidates; the \Bs candidate flight distance and
impact parameter with respect to the PV;
the $\psi(2S)$ candidate vertex $\chi^2$; the $\chi^2_{\rm IP}$ of
the kaon and muon candidates (defined as the change in $\chi^2$ of the PV fit
when reconstructed with and without the
considered particle) and the muon identification probabilities. The optimal working
point for the BDT is determined using a figure of merit that optimises the
statistical power of the selected data sample for the analysis of $\phis$
by taking account of the  number of signal and background candidates,
as well as the decay-time resolution and flavour-tagging power of each candidate.

\begin{figure}[t]
\centering
  \includegraphics[width=0.6\textwidth]{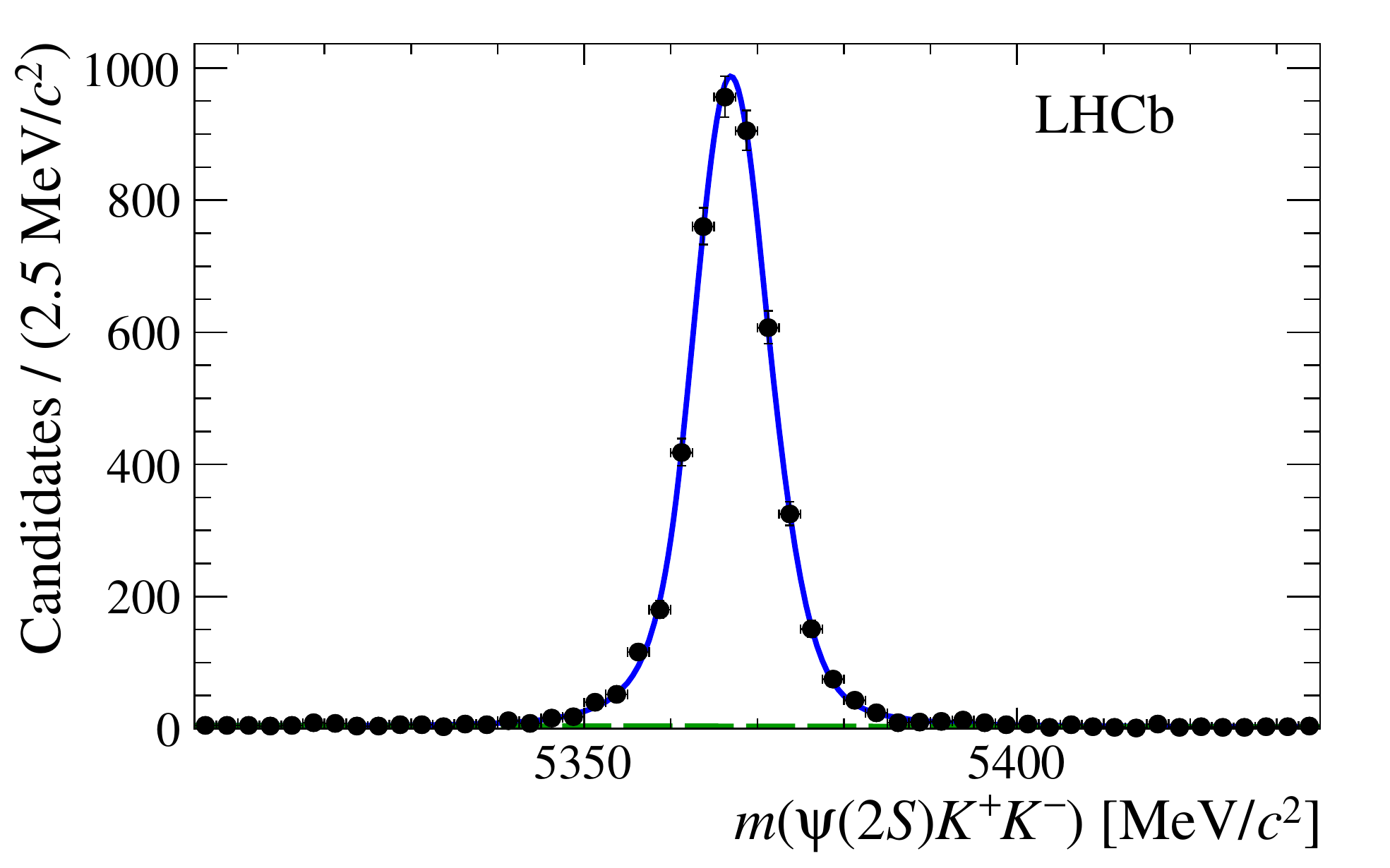}
  \caption{\small Distribution of $m(\psi(2S) \Kp\Km)$
  for the selected  $\Bs\to\psi(2S)\phi$ candidates.
  The total fit model is shown by the solid blue line,
  which is composed of a sum of two Crystal Ball functions
  for the signal and an exponential function for the background (long-dashed green line).}
  \label{fig:bs_mass}
\end{figure}

Figure~\ref{fig:bs_mass} shows the distribution of $m(\psi(2S) \Kp\Km)$
for the selected  $\Bs\to\psi(2S)\phi$ candidates. An extended maximum likelihood fit is made to the
unbinned $m(\psi(2S) \Kp\Km)$ distribution, where the signal component is described by
the sum of two Crystal Ball~\cite{Skwarnicki:1986xj} functions and the small combinatorial
background by an exponential function.
All parameters are left free in the fit, including the yields of the signal and background components.
This fit gives a yield of $4695 \pm 71$ signal candidates  and
$174 \pm 10$ background candidates in the range $m(\psi(2S) \Kp\Km) \in [5310, 5430]\mevcc$.
It is used to assign per-candidate weights (sWeights) via the sPlot technique~\cite{Pivk:2004ty},
which are used to subtract the background contribution in the maximum likelihood fit
described in Section~\ref{sec:Results}.

\section{Detector resolution and efficiency}
\label{sec:Resolution}

The resolution on the measured decay time is determined
with the same method as
described in Refs.~\cite{LHCb-PAPER-2014-059,LHCb-PAPER-2013-002} by
using a large sample of prompt $\jpsi\Kp\Km$ combinations produced directly in the $pp$ interactions. These
events are selected using prompt ${\jpsi \to \mu^+\mu^-}$ decays via a prescaled
trigger that does not impose any requirements on the separation of the \jpsi
from the PV. The \jpsi candidates are combined with oppositely charged tracks that are identified
as kaons, using a similar selection as for the signal decay. The
resolution model, $R(t-t^\prime)$, is the sum of two Gaussian distributions with per-event
widths.
These widths are calibrated by using a
maximum likelihood fit to the unbinned decay time and decay-time uncertainty distributions
of the prompt $\jpsi\Kp\Km$ combinations, using a model composed of the sum
of a $\delta$ function for the prompt component and two exponential functions for long-lived backgrounds,
all of which are convolved with the resolution function.
A third Gaussian distribution is added to the total fit function to account
for the small ($<1\%$) fraction of decays that are associated to the wrong PV.
The average effective resolution is $46.6 \pm 1.0$\fs.
Simulated $\decay{\Bs}{\jpsi \Kp \Km}$ and
$\decay{\Bs}{\psi(2S) \Kp \Km}$ events show
no significant difference in the effective decay-time resolution between the two decay modes.

The reconstruction efficiency is not constant as a function of decay time due to
displacement requirements made on signal tracks in the trigger and event selection. The
efficiency is determined using the control channel $\Bd \to \psi(2S)K^{*}(892)^0$,
with $K^{*}(892)^0 \to \Kp\pi^-$, which is assumed to have
a purely exponential decay-time distribution. It is defined as
\begin{equation}
	\varepsilon_{\rm data}^{\Bs}(t) = \varepsilon_{\rm data}^{\Bd}(t) \times \frac{\varepsilon_{\rm sim}^{\Bs}(t)}{\varepsilon_{\rm sim}^{\Bd}(t)},
\end{equation}
where $\varepsilon_{\rm data}^{\Bd}(t)$ is the efficiency of the control channel and
$\varepsilon_{\rm sim}^{\Bs}(t)/\varepsilon_{\rm sim}^{\Bd}(t)$ is the ratio of efficiencies of the
simulated signal and control modes after the full trigger and selection chain has been applied.
This correction accounts for the small differences in the lifetime and kinematics between
the signal and control modes.

The $\Bd \to \psi(2S)K^{*}(892)^0$ decay is selected using a similar trigger, preselection
and the same BDT training and working point as used for the signal (with appropriate changes for kaon to pion).
Backgrounds from the misidentification of final-state particles
from other decays such as $\Bs\to\psi(2S)\phi$
and $\Lb\to\psi(2S)pK^-$ are negligible.
Similarly, possible backgrounds from $\B_{(s)}^0\to\psi(2S)\pi^+\pi^-$ decays where a
pion is misidentified as a kaon, and $\Bu\to \psi(2S)\Kp$ decays combined with an
additional random pion, are negligible.

\begin{figure}[t]
\centering{
    \includegraphics*[width=0.6\textwidth]{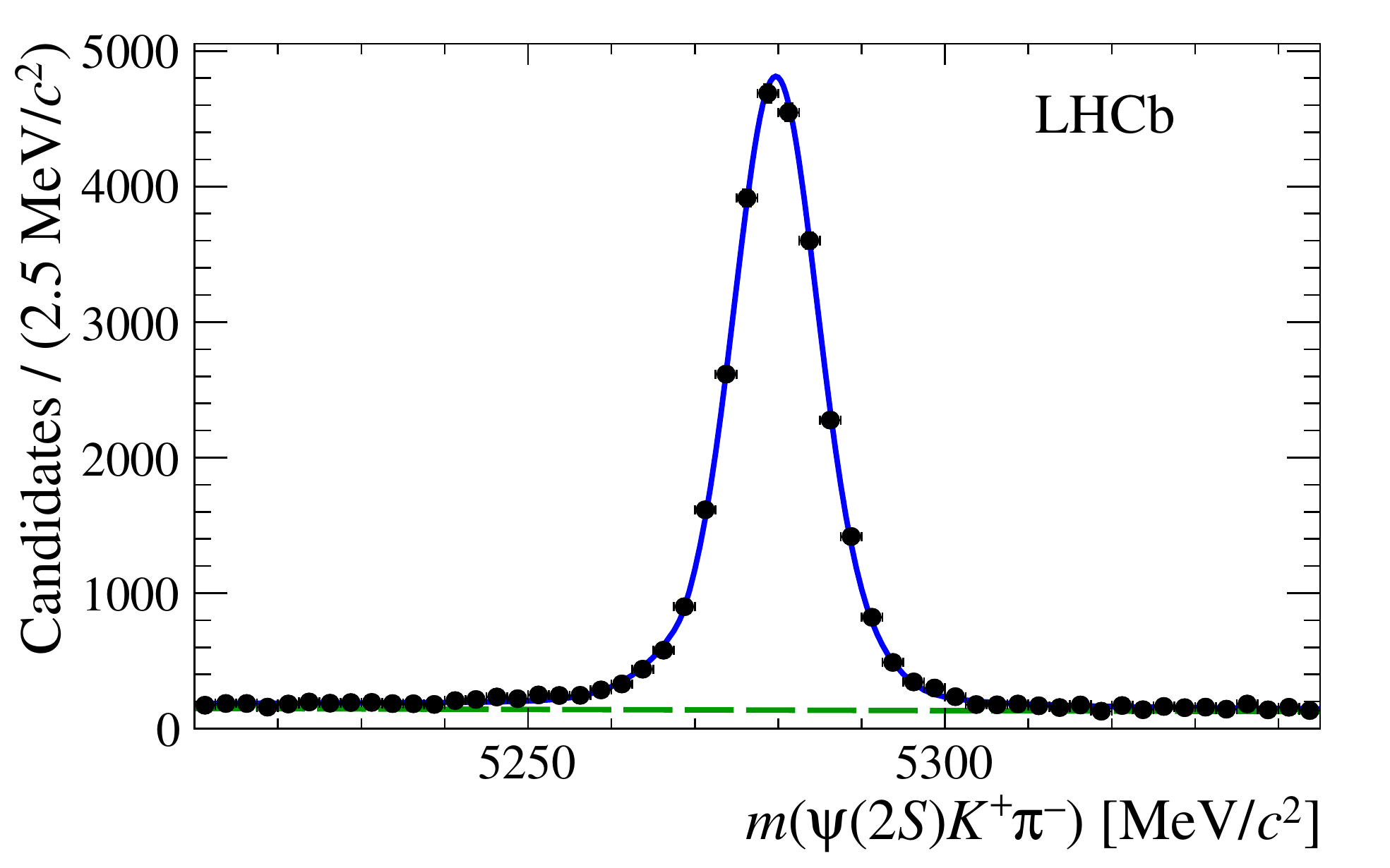}
    }
  \caption{\small Distribution of $m(\psi(2S) \Kp\pi^-)$ of the selected  $\Bd\to\psi(2S)K^{*}(892)^{0}$ candidates.
  The total fit model is shown by the solid blue line,
  which is composed of a sum of two Crystal Ball functions
  for the signal and an exponential function for the background (long-dashed green line).}
  \label{fig:bd_mass}
\end{figure}

\begin{figure}[t]
\centering{
    \includegraphics*[width=0.6\textwidth]{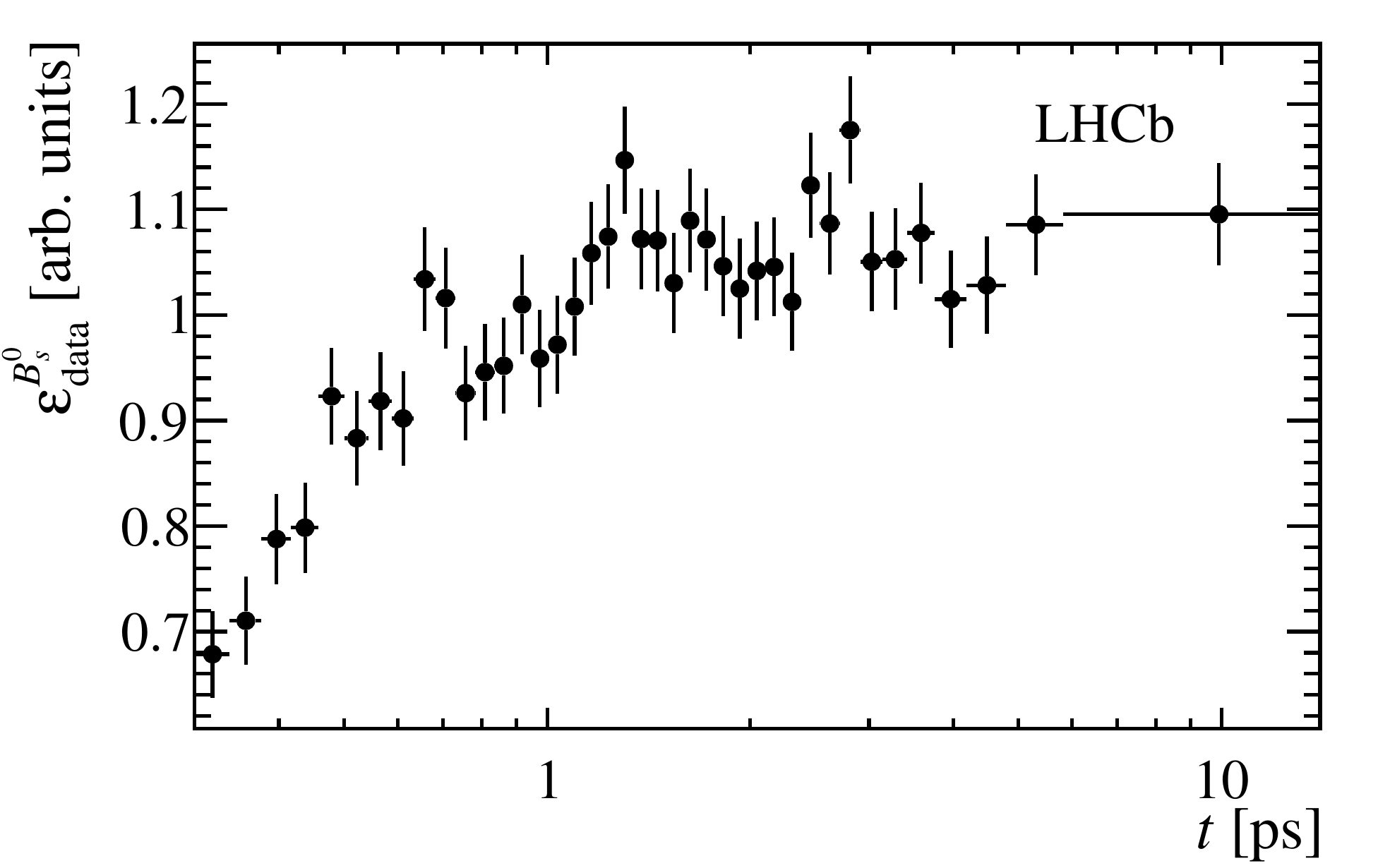}
    }
  \caption{\small Decay-time efficiency $\varepsilon_{\rm data}^{\Bs}(t)$ in arbitrary units. }
  \label{fig:bs_data}
\end{figure}

The $\psi(2S)K^+\pi^-$
invariant mass distribution is shown in Fig.~\ref{fig:bd_mass} along with the result of a fit
composed of the sum of two Crystal Ball (CB) functions for the signal and an exponential function
for the background. The tail parameters and relative fraction of the two CB functions
are fixed to values obtained from a fit to simulated $\Bd \to \psi(2S)K^{*}(892)^0$ decays.
The core widths and common mean of the CB functions are free in the fit
and the \Bd yield is found to be $28\,676\pm195$. The efficiency is defined as
$\varepsilon_{\rm data}^{\Bd}(t) = N_{\rm data}^{\Bd}(t)/N_{\rm gen}^{\Bd}(t)$
where $N_{\rm data}^{\Bd}(t)$ is the number of signal $\Bd \to \psi(2S)K^{*}(892)^0$ decays
in a given bin of decay time and
$N_{\rm gen}^{\Bd}(t)$ is the number of events generated from an
exponential distribution with lifetime $\tau_{\Bd} = 1.520 \pm 0.004\ps$~\cite{PDG2014}.
The exponential distribution is convolved with a double Gaussian resolution model,
the parameters of which are determined
from a fit to the decay time distribution of prompt $\jpsi\Kp\pi^-$ combinations.
In total $10^7$ events are generated.
The sPlot~\cite{Pivk:2004ty} technique with $m(\psi(2S)K^+\pi^-)$ as discriminating variable
is used to determine $N_{\rm data}^{\Bd}(t)$.
The analysis is not sensitive to the absolute scale of the efficiency.
The final decay-time efficiency for the $\decay{\Bs}{\psi(2S) \phi}$ signal is
shown in Fig.~\ref{fig:bs_data}.
It is relatively uniform at high values of decay time but
decreases at low decay times due to selection requirements placed on the
track $\chi^2_{\rm IP}$ variables.

The efficiency as a function of the $\Bs\to\psi(2S)\phi$ helicity angles is not
uniform due to the forward geometry of the LHCb detector and the
requirements imposed on the final-state
particle momenta. The three-dimensional efficiency, $\varepsilon(\Omega)$,
is determined with the same technique as used in Ref.~\cite{LHCb-PAPER-2013-002}
using simulated events that are subjected to
the same trigger and selection criteria as the data.
The relative efficiencies vary by up to 20\%, dominated
by the dependence on $\cos\theta_\mu$.

\section{Flavour tagging}
\label{sec:tagging}

\DeclareRobustCommand{\optbar}[1]{\shortstack{{\miniscule (\rule[.5ex]{1.25em}{.18mm})}
  \\ [-.7ex] $#1$}}
\def\BsorBsb    {\ensuremath{\optbar{B}{}^0_s}\xspace}

The \Bs candidate flavour at production is determined by two independent classes of
flavour tagging algorithms, the opposite-side (OS) taggers~\cite{LHCb-PAPER-2011-027} and the same-side kaon (SSK)
tagger~\cite{LHCb-PAPER-2015-056}, which exploit specific features of the production of $\bquark \bquarkbar$
quark pairs in $\proton\proton$ collisions, and their subsequent hadronisation.
Each tagging algorithm gives a tag decision and a mistag probability.
The tag decision, $\mathfrak{q}$, takes values $+1$, $-1$, or $0$, if the signal meson is tagged as $\Bs$, $\Bsb$, or is untagged,
respectively. The fraction of events in the sample with a nonzero tagging decision gives
the efficiency of the tagger, $\varepsilon_{\rm tag}$.
The mistag probability, $\eta$, is estimated event-by-event, and
represents the probability that the algorithm assigns a wrong tag decision to the event; it
is calibrated using data samples of several flavour-specific $\Bd$, $\Bu$ and $B_{s2}^{*0}$ decays to obtain
the corrected mistag probability, $\optbar{\omega}$, for an initial
flavour \BsorBsb meson. A linear relationship between $\eta$
and $\optbar{\omega}$ is used for the calibration. The effective tagging power is
given by $\varepsilon_{\rm tag}(1-2\omega)^2$
and for the
combined taggers in the $\Bs\to\psi(2S)\phi$ signal sample is $(3.88 \pm 0.13 \pm 0.12)\%$,
where the first uncertainty is statistical and the second systematic.

\section{Maximum likelihood fit}
\label{sec:Results}

The physics parameters are determined by a weighted maximum likelihood fit of
a signal-only probability density function (PDF)
to the four-dimensional distribution of \mbox{$\Bs\to\psi(2S)\phi$} decay time and helicity angles.
The negative log-likelihood function to be minimised is given by
\begin{equation}
-\ln {\cal L} = -\alpha \sum_{{\rm events}\ i} W_i \ln\, {\cal P},
\label{eq:all}
\end{equation}
where $W_i$ are the sWeights computed using $m(\psi(2S)\Kp\Km)$ as the discriminating variable and
the factor $\alpha = \sum W_i/\sum W_i^2$ is necessary to obtain the
correct parameter uncertainties from the Hessian of the negative log-likelihood.
The PDF, ${\cal P} = {\cal S}/\int {\cal S}\,\deriv t \;\deriv\Omega$,
is obtained from
\begin{equation}
{\cal S}(t, \Omega, \mathfrak{q}^{\rm OS}, \mathfrak{q}^{\rm SSK}|\eta^{\rm OS}, \eta^{\rm SSK})
= {\cal X}(t^\prime, \Omega, \mathfrak{q}^{\rm OS}, \mathfrak{q}^{\rm SSK}|\eta^{\rm OS}, \eta^{\rm SSK})
\otimes R(t-t^\prime) \times \varepsilon_{\rm data}^{\Bs}(t),
\label{eq:pdf}
\end{equation}
where
\begin{equation}
\begin{aligned}
{\cal X}(t, \Omega, \mathfrak{q}^{\rm OS}, \mathfrak{q}^{\rm SSK}|\eta^{\rm OS}, \eta^{\rm SSK}) = &\ \left( 1+ \mathfrak{q}^{\rm OS}(1-2\mistag^{\rm OS})\right)\left(1+\mathfrak{q}^{\rm SSK}(1-2\mistag^{\rm SSK})\right) X(t,\Omega) + \\
&\ \left( 1 - \mathfrak{q}^{\rm OS}(1-2\bar{\mistag}^{\rm OS})\right)\left(1-\mathfrak{q}^{\rm SSK}(1-2\bar{\mistag}^{\rm SSK})\right) \overline{X}(t,\Omega),
\end{aligned}
\label{eq:decayrate}
\end{equation}
which allows for the inclusion of information from both tagging algorithms in the computation of the
decay rate. The function $X(t,\Omega)$ is defined in Eq.~\ref{eq:Eqbsrate} and $\overline{X}(t,\Omega)$
is the corresponding function for \Bsb decays.
As in Ref.~\cite{LHCb-PAPER-2013-002}, the angular efficiency is included in the normalisation of the PDF
via ten integrals, $I_k = \int \deriv\Omega\, \varepsilon(\Omega) f_k(\Omega)$,  which are
calculated using simulated events.
In contrast to Refs.~\cite{LHCb-PAPER-2014-059,LHCb-PAPER-2013-002},
the fit is performed in a single bin of $m(\Kp\Km)$, within $12\mevcc$ of the known $\phi$ mass.

\begin{figure}[t]
  \begin{center}
    \includegraphics[width=0.49\linewidth]{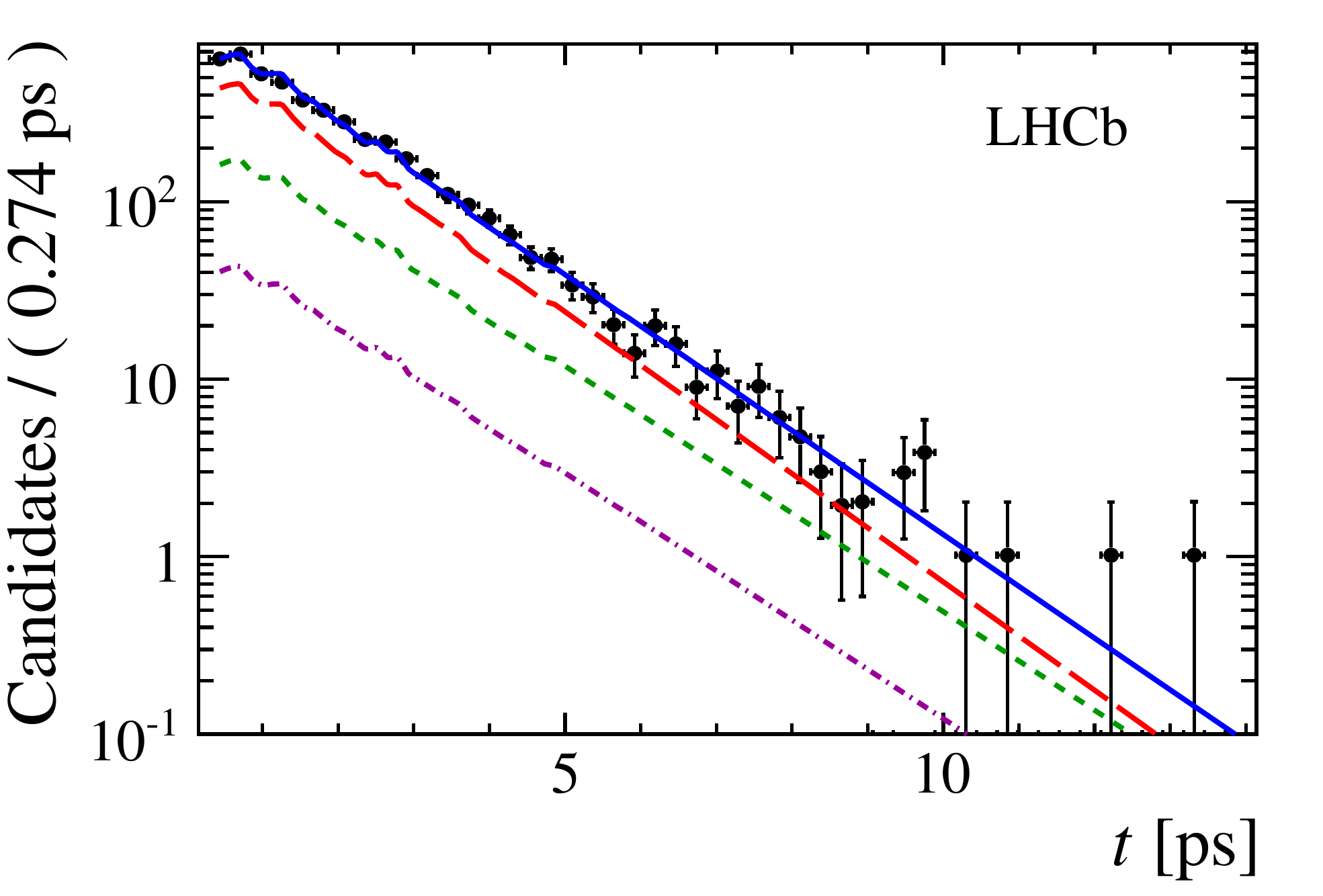}
    \includegraphics[width=0.49\linewidth]{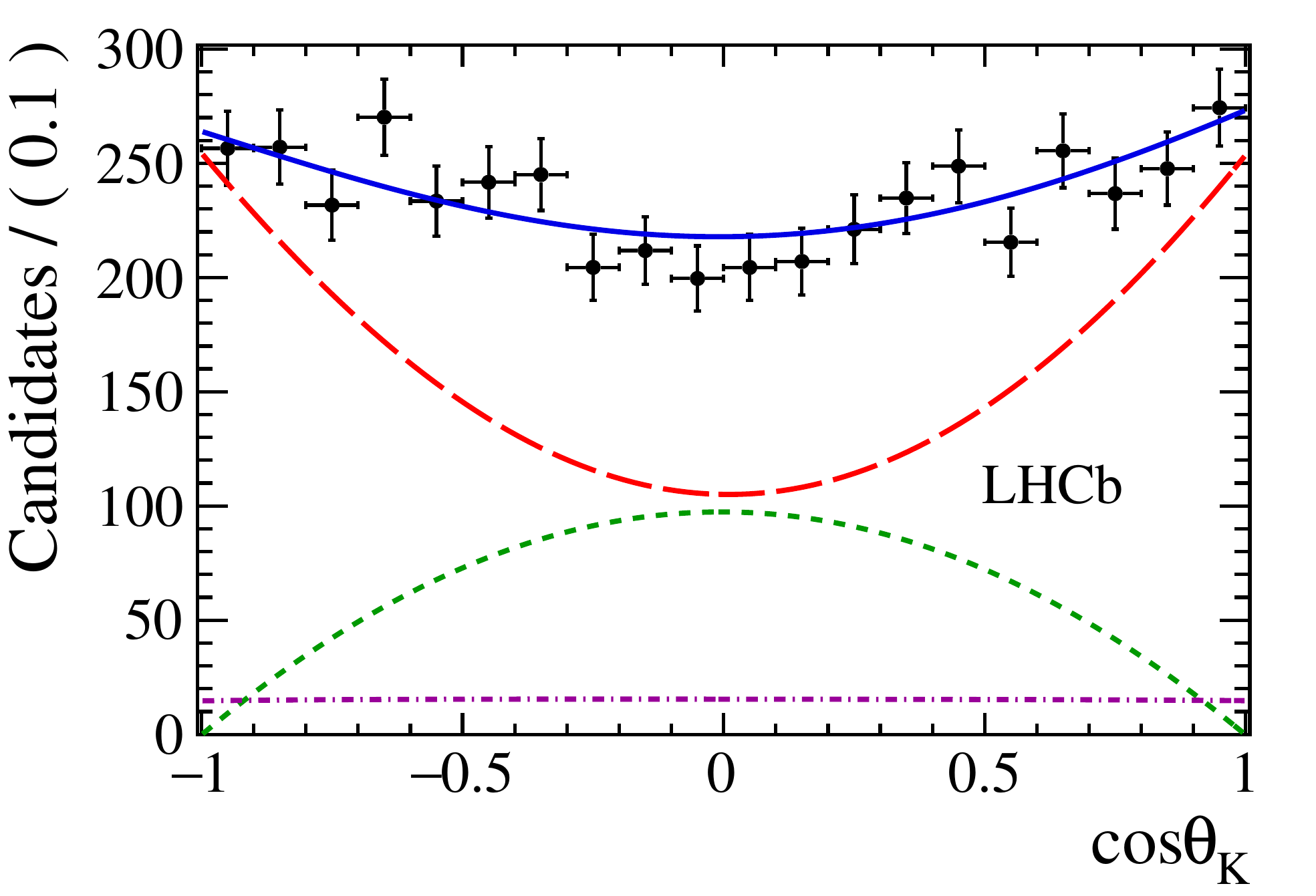}\\
    \includegraphics[width=0.49\linewidth]{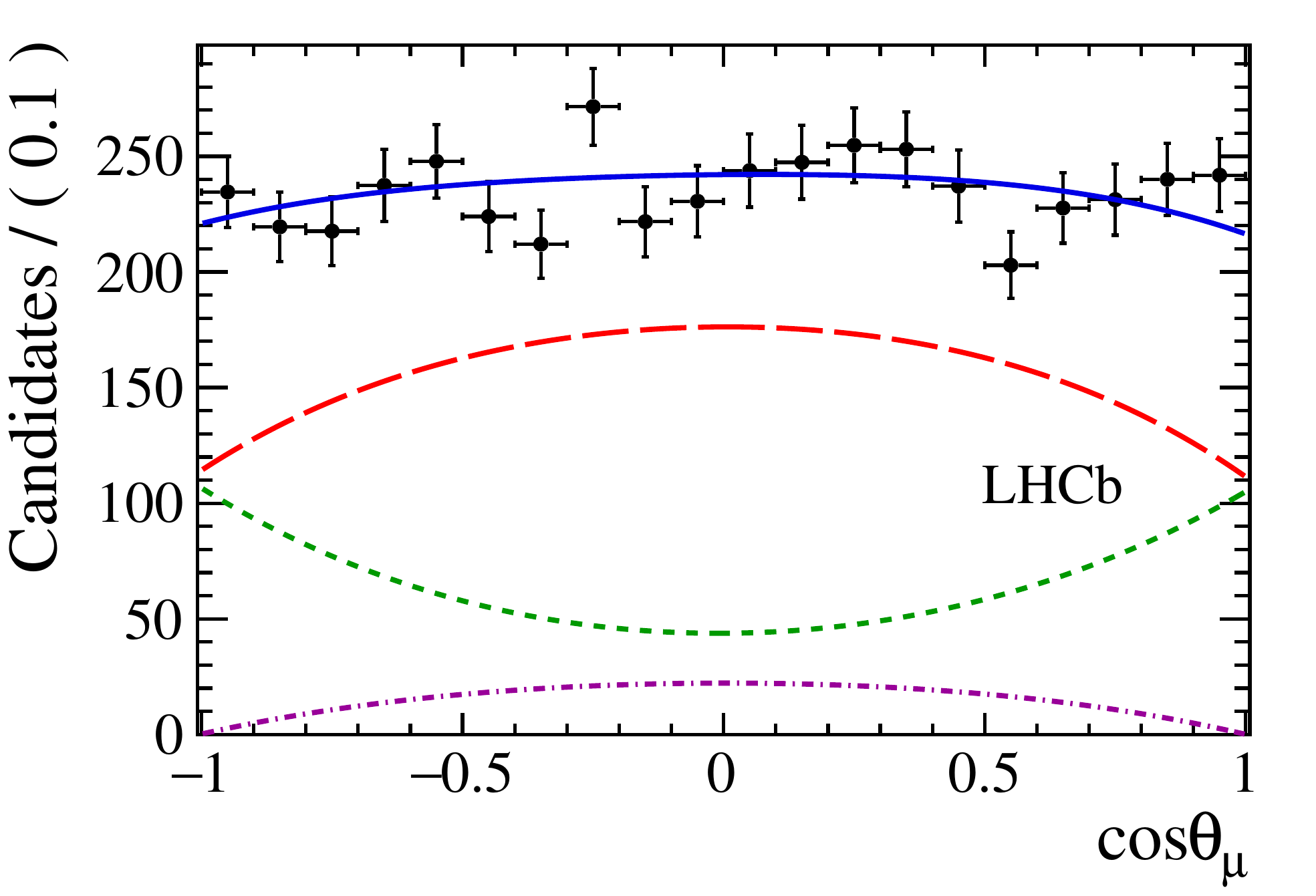}
    \includegraphics[width=0.49\linewidth]{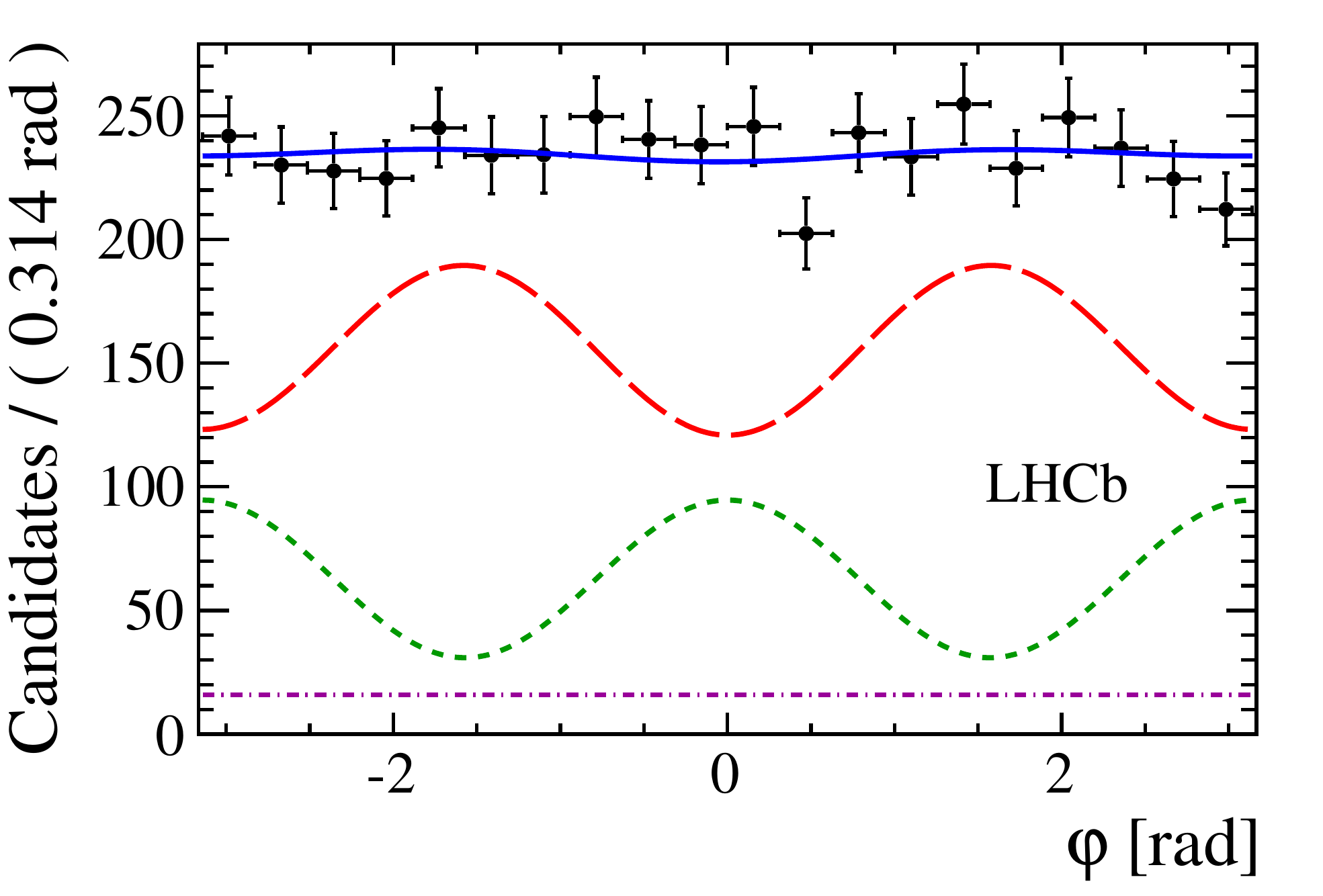}
  \end{center}
  \caption{
    \small
    Decay-time and helicity-angle distributions for $\Bs\to\psi(2S)\phi$ decays (data points)
    with the one-dimensional projections of the fitted PDF.
    The solid blue line shows the total signal contribution, which is composed of \CP-even (long-dashed red),
     \CP-odd (short-dashed green) and $S$-wave (dash-dotted purple) contributions.}
  \label{fig:res:projthetak}
\end{figure}

In the fit, Gaussian constraints are applied to the \Bs mixing frequency
\mbox{$\Delta m_s = 17.757 \pm 0.021\invps$}~\cite{HFAG}
and the tagging calibration parameters. The fitting procedure has been validated using pseudoexperiments
and simulated $\Bs\to\psi(2S)\phi$ decays.
Due to the symmetry in the PDF there is a two-fold ambiguity in the solutions for
$\phis$ and $\DGs$; the solution with positive $\DGs$ is used~\cite{LHCb-PAPER-2011-028}.
The results of the fit to the data are shown in
Tables~\ref{table:results_with_systematics} and~\ref{tab:correlations} while the projections
of the fit onto the data are shown in Fig.~\ref{fig:res:projthetak}.
The results are consistent with previous
measurements of these parameters~\cite{Abazov:2011ry,CDF:2011af,LHCb-PAPER-2014-059,Aad:2016tdj,Khachatryan:2015nza},
and the SM predictions for $\phis$ and $\DGs$~\cite{Charles:2004jd,Lenz:2006hd,Artuso:2015swg}.
They show no evidence of \CP violation in the interference between \Bs meson mixing and decay,
nor for direct \CP violation in $\Bs \to \psitwos \phi$ decays as the
parameter $|\lambda|$ is consistent with unity.
The likelihood profile for $\delta_\parallel$ is not parabolic
and the 95\% confidence level range is $[2.4, 3.9]$\rad.

\begin{figure}[t]
  \begin{center}
    \includegraphics[width=0.6\linewidth]{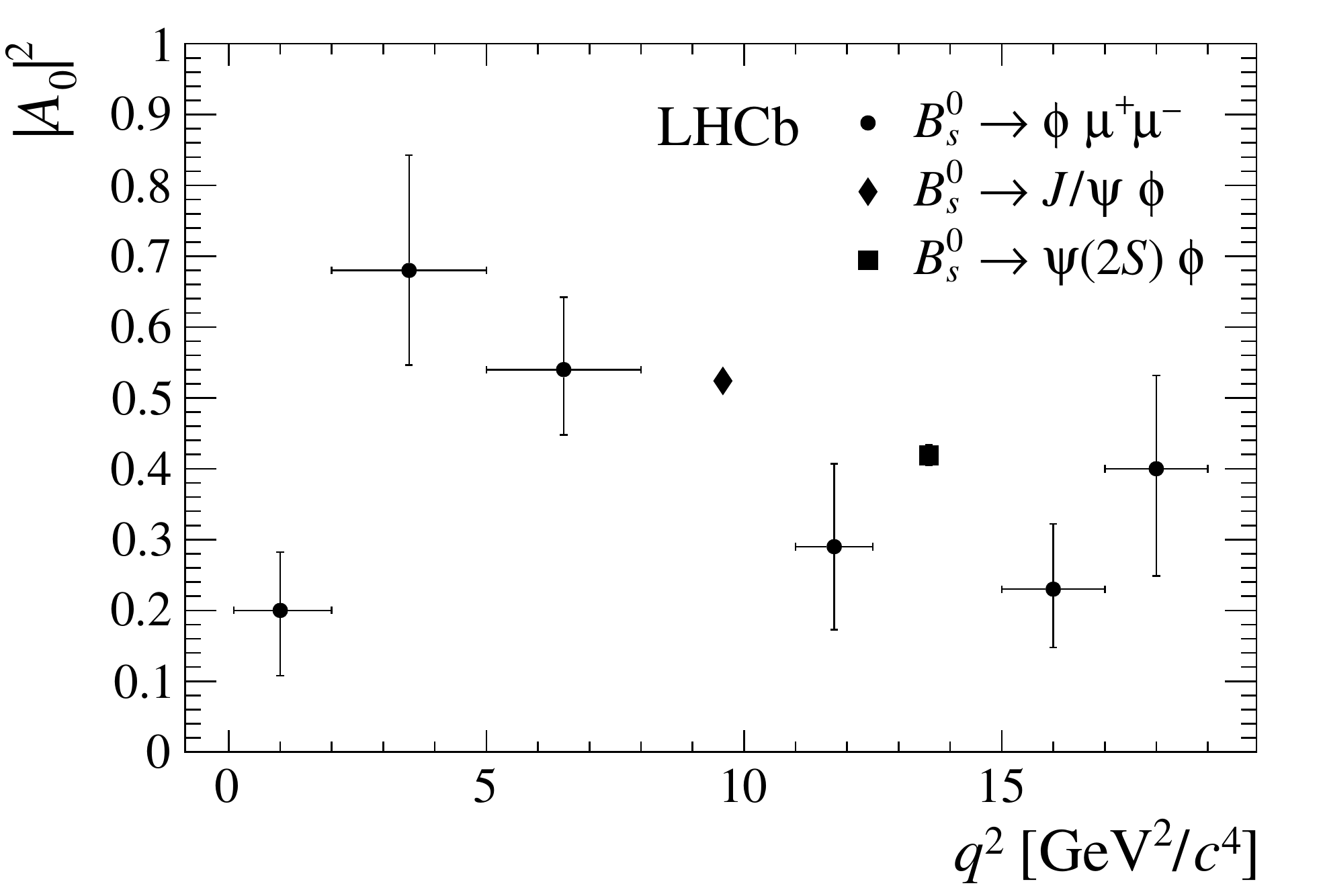}
  \end{center}
  \caption{
    \small
    $|A_0|^2$ as a function of the invariant mass squared of the dimuon system,
    $q^2$. Data points are taken from Ref.~\cite{LHCb-PAPER-2015-023}
    ($\Bs\to\phi\mu^+\mu^-$, circles), Ref.~\cite{LHCb-PAPER-2014-059} ($\Bs\to\jpsi\phi$, diamond)
    and this paper (square).}
  \label{fig:FL}
\end{figure}

Figure~\ref{fig:FL} shows values of $F_{\rm L} \equiv |A_0|^2$, the fraction
of longitudinal polarisation, for $\Bs\to\phi \mu^+\mu^-$~\cite{LHCb-PAPER-2015-023},
$\Bs\to \jpsi\phi$~\cite{LHCb-PAPER-2014-059} and $\Bs\to \psi(2S)\phi$
final states as a function of the invariant mass squared of the dimuon system, $q^2$.
The precise measurement of $F_{\rm L}$ from $\Bs\to \jpsi\phi$ at $q^2 = 9.6$ GeV$^2$/$c^4$
is now joined by the precise measurement from this paper at $q^2 = 13.6$ GeV$^2$/$c^4$,
demonstrating a clear decrease with $q^2$ towards the value of 1/3, as predicted by Ref.~\cite{Hiller:2013cza}.

\begin{table}[t]
\caption{\small Results of the maximum likelihood fit to the selected $\Bs\to\psi(2S)\phi$ candidates
including all acceptance and resolution effects.
The first uncertainty is statistical and the second is systematic, which will be discussed in
Section~\ref{sec:Systematics}.}
\begin{center}
\begin{tabular}{@{}lr@{}}
Parameter & Value\hspace{1.3cm} \\               \hline
$\Gamma_{s}$ [$\invps$] 		&    $0.668 \pm    0.011 \pm 0.006$\rule[-2mm]{0mm}{6mm}\\
$\Delta\Gamma_{s}$ [$\invps$] &    $0.066^{+0.041}_{-0.044} \pm 0.007$\rule[-2mm]{0mm}{6mm} \\
$|A_{\hspace{-1pt}\perp}|^{2}$ 	&    $0.264^{+0.024}_{-0.023} \pm 0.002$\rule[-2mm]{0mm}{6mm}\\
$|A_0|^2$ 				&    $0.422 \pm 0.014 \pm 0.003$\rule[-2mm]{0mm}{6mm} \\
$\delta_\parallel$  [rad]		&    $3.67^{+0.13}_{-0.18} \pm 0.03$\rule[-2mm]{0mm}{6mm}\\
$\delta_\perp$	 [rad]			&    $3.29^{+0.43}_{-0.39} \pm 0.04$\rule[-2mm]{0mm}{6mm}\\
$\phis$ [rad] 				&    $0.23^{+0.29}_{-0.28} \pm 0.02$\rule[-2mm]{0mm}{6mm}\\
$|\lambda|$ 				&    $1.045^{+0.069}_{-0.050} \pm 0.007$\rule[-2mm]{0mm}{6mm} \\
$F_{S}$ 					&    $0.061^{+0.026}_{-0.025} \pm 0.007$\rule[-2mm]{0mm}{6mm}\\
$\delta_{S}$ [rad] 			&    $0.03 \pm     0.14 \pm 0.02$\rule[-2mm]{0mm}{6mm}\\
           \hline
\end{tabular}
\label{table:results_with_systematics}
\end{center}
\end{table}

\begin{table}
\caption{\small Correlation matrix of statistical uncertainties.}
\begin{center}
\begin{tabular}{@{}l|r r r r r r r r r r @{}}
& $\Gamma_s$ & $\Delta\Gamma_s$ & $|A_{\hspace{-1pt}\perp}|^{2}$ & $|A_0|^2$ & $\delta_\parallel$ & $\delta_\perp$ & $F_{S}$ & $\delta_{S}$ & $\phi_s$ & $|\lambda|$ \\  \hline
$\Gamma_s$ & 1.00 & $-$0.40 & 0.35 & $-$0.27 & $-$0.08 & $-$0.02 & 0.15 & 0.02  & 0.02 & $-$0.04  \\
$\Delta\Gamma_s$ &  & 1.00 & $-$0.66 & 0.60 & 0.02 & $-$0.04 & $-$0.10 & $-$0.02  & 0.19 & 0.03  \\
$|A_{\hspace{-1pt}\perp}|^{2}$ &  &  & 1.00 & $-$0.54 & $-$0.31 & $-$0.05 & 0.08 & 0.03  & $-$0.02 & $-$0.02  \\
$|A_0|^2$ &  &  &  & 1.00 & 0.05 & $-$0.02 & $-$0.15 & $-$0.02  & 0.07 & 0.03  \\
$\delta_\parallel$ &  &  &  &  & 1.00 & 0.26 & $-$0.26 & $-$0.01  & 0.00 & 0.08 \\
$\delta_\perp$ &  &  &  &  &  & 1.00 & $-$0.21 & $-$0.25  & $-$0.06 & 0.59  \\
$F_{S}$ &  &  &  &  &  &  & 1.00 & 0.02  & 0.05 & $-$0.25  \\
$\delta_{S}$ &  &  &  &  &  &  &  & 1.00  & 0.07 & $-$0.09  \\
$\phi_s$ &  &  &  &  &  &  &  &  &  1.00 & 0.04  \\
$|\lambda|$ &  &  &  &  &  &  &  &  &  & 1.00  \\

\end{tabular}
\label{tab:correlations}
\end{center}
\end{table}

\section{Systematic uncertainties}
\label{sec:Systematics}

\def\delone{\delta_1}
\def\deltwo{\delta_2}
\def\delpar{\delta_\parallel}
\def\delperp{\delta_\perp}
\def\delzero{\delta_0}
\def\dels{\delta_{\rm S}}
\def\delsperp{\dels - \delperp}

\def\apar{A_{\|}}
\def\aperp{A_{\perp}}
\def\aparsq{|A_{\|}|^2}
\def\aperpsq{|A_{\perp}|^2}
\def\assq{|A_{s}|^2}
\def\azero{A_{0}(t)}
\def\azerosq{|A_{0}|^2}

\def\aparsqo{|A_{\|}(0)|^2}
\def\aperpsqo{|A_{\perp}(0)|^2}
\def\assqo{|A_{s}(0)|^2}
\def\azerosqo{|A_{0}(0)|^2}
\def\fS{F_{\text{S}}}
\def\maglambda{|\lambda|}

{
\begin{table}[tb]
\setlength{\tabcolsep}{5pt}
  \caption{\small Summary of statistical and systematic uncertainties.
  Fields containing a dash (--) correspond to systematic uncertainties that are negligible.}
\begin{scriptsize}
  \begin{center}\footnotesize
    \begin{tabular}{lccccccccccc}
      Source      & $\Gs$         & $\DGs$        & $\aperpsq$    & $\azerosq$    & $\delpar$     &$\delperp$     & $\phi_s$      & $\maglambda$ & $ F_{S} $&$ \delta_{S}$         \\
                                             & [ps$^{-1}$]   &[ps$^{-1}$]    &               &               &  [rad]        &  [rad]        & [rad]         &        & & [rad]     \\[0.5ex] \hline
      \rule[3.5mm]{0mm}{0mm}Stat. uncertainty
      					   & 0.011       & $^{+0.041}_{-0.044}$        & $^{+0.024}_{-0.023}$        & $0.014$        & $ ^{+0.13}_{-0.18}$ & $ ^{+0.43}_{-0.39}$            & $^{+0.29}_{-0.28}$  &     $^{+0.069}_{-0.050}$ & $^{+0.026}_{-0.025}$ &  $0.14$ \\[0.1ex]
      \hline
      Mass factorisation	                 & 0.003 &    0.002     &  0.001 &  0.001   &   0.02     &   --      &     0.01     &     0.001 & 0.003 & 0.01              \\
      Mass model                			& 0.001 & 0.001     & --  & --   &  --     & --        &   --    & 0.001     & --&  --       \\
      Angular eff. (stat.)                        	&  --    &  0.001      & 0.001  &   0.002    &  0.02       &   0.03     & 0.01   & 0.006  & 0.005& 0.02   \\
      Angular resolution                       	&  --   &    --         &    0.001     &  --       &    0.01        &    0.01      &     --       & --    & -- &   --                    \\
      Time resolution                         	&  --   &  0.001   &  --   &  --  & --     & 0.02  &  0.02  & 0.002 & 0.002 &  --      \\
      Time resolution  (stat.)                	&   --    &  --   &  --    & --     & --    & 0.02   & -- & 0.002 &--  & --   \\
      Time eff. (stat.)       		 	& 0.005   &  0.003   &  0.001   & 0.001    &  --   & --      &  -- &-- &   0.002 & --   \\
      Time eff. (mass model)	& 0.001  &  0.001   & --    &  --   &  --   &  --     &  -- & -- &  --  & --   \\
      Time eff. ($\tau_{\Bd}$) & 0.002 & -- & -- & -- & -- & -- & -- & -- & -- & -- \\
      $B_c^+$ feed-down                   &  0.001   &      --       &    --         &       --      &     --        &        --             &  --           &         --       &  --           &         --                  \\
      Fit bias                                  	&  0.001   &    0.006         &    --     &   0.001      &    0.01        &    --      &     --       & --    & 0.003 &   --                    \\
      \hline
      Quad. sum of syst. 		   	&  0.006   &   0.007  & 0.002  &  0.003 & 0.03       &   0.04     &  0.02      &  0.007 & 0.007 & 0.02       \\
      Total uncertainties                       	&  0.013   &   $^{+0.042}_{-0.045}$  & $^{+0.024}_{-0.023}$  &  $0.014$ & $ ^{+0.13}_{-0.18}$       &   $ ^{+0.43}_{-0.39}$     &  $^{+0.29}_{-0.28}$      &  $^{+0.069}_{-0.050}$ & $^{+0.027}_{-0.026}$ & 0.14       \\

      \hline
      \end{tabular}
  \end{center}  \label{table:finalsystematicssummary}
\end{scriptsize}
\end{table}
}

Systematic uncertainties for each of the measured parameters are reported in \autoref{table:finalsystematicssummary}.
They are evaluated by observing the change in physics
parameters after repeating the likelihood fit with a modified model assumption, or
by generating pseudoexperiments in case of uncertainties originating from the limited
size of a calibration sample.
In general the sum in quadrature of the different sources of systematic uncertainty is less than 20\% of the statistical
uncertainty, except for $\Gamma_s$ where it is close to 60\%.

Repeating the fit to $m(\psi(2S)\Kp\Km)$ in bins of the decay time and helicity angles shows that the mass
resolution depends upon $\cos\theta_\mu$. This breaks the assumption  that $m(\psi(2S)\Kp\Km)$ is
uncorrelated with the observables of interest, which is implicitly made by the use of
weights from the sPlot technique.
The effect of this correlation is quantified by repeating the four-dimensional likelihood
fit for different sets of signal weights computed from fits to $m(\psi(2S)\Kp\Km)$ in bins of
$\cos\theta_\mu$. The largest variation in each physics parameter is assigned a systematic uncertainty.
The mass model is tested by computing a new set of sWeights, using a Student's $t$-function to describe the
signal component of the $m(\psi (2S) \Kp\Km)$ distribution.

The statistical uncertainty on the angular efficiency is propagated by repeating the fit using
new sets of the ten integrals, $I_k$, systematically varied according to their covariance matrix.
The effect of assuming perfect angular resolution in the likelihood fit is studied using pseudoexperiments.
There is a small effect on the polarisation amplitudes and strong phases while all
other parameters are unaffected.

The decay-time resolution is studied by generating pseudoexperiments using the nominal
double Gaussian model and subsequently fitting them using a single Gaussian model,
the parameters of which have been calibrated on the prompt $\jpsi\Kp\Km$ sample.
In addition, the nominal model parameters are varied within their statistical uncertainties and the
fit repeated.

The decay-time efficiency introduces a systematic uncertainty from three different sources.
First, the contribution due to the statistical error on the determination of the decay-time efficiency from
the control channel is determined by repeating the fit multiple times after randomly varying the
parameters of the time efficiency within their statistical uncertainties.
The statistical uncertainty is dominated by the size of the $\Bd \to \psi(2S)K^{*}(892)^0$ control sample.
Second, a Student's $t$-function is used as an alternative mass model
for the $m(\psi(2S)\Kp\pi^-)$ distribution
and a new decay-time efficiency function is produced.
Finally, the efficiency  function is recomputed with the lifetime of the \Bd modified by $\pm1\sigma$.
In all cases the difference in fit results arising from the use of the new efficiency function is
taken as a systematic uncertainty.
The sensitivity to the BDT selection
is studied by adjusting the working point around the optimal position equally for both signal and control channel, and also differently
for each channel in order to make the ratio $\varepsilon_{\rm sim}^{\Bs}(t)/\varepsilon_{\rm sim}^{\Bd}(t)$ uniform.
The efficiency is recomputed in each case and the fit repeated. No significant change in the
physics parameters is observed.

A small fraction of $\Bs\to \psi(2S)\phi$ signal candidates comes from the decay of $B_c^+$ mesons,
causing an average positive shift in the reconstructed decay time of the $\Bs$ meson.
This fraction was estimated as $0.8\%$ in Ref.~\cite{LHCb-PAPER-2014-059} and pseudoexperiments were
used to assess the impact of ignoring such a contribution.
Only $\Gamma_s$ was affected, with a bias on its central value of $(+20\pm6)\%$
of its statistical uncertainty.
The assumption is made that the ratio of efficiencies
for selecting $\Bs\to \psi(2S)\phi$ decays either promptly or via the
decay of $B_c^+$ mesons is the same as that for $\Bs\to \jpsi\phi$ decays.
This leads to a bias of $+0.002 \pm 0.001\invps$ in $\Gamma_s$.
The central value of $\Gamma_s$ is therefore reduced by 0.002\invps
and a systematic uncertainty of 0.001\invps is assigned.

A test for a possible bias in the fit procedure is performed by generating and fitting
many simulated pseudoexperiments of equivalent size to the data sample.
The resulting biases are small and those that are not compatible with zero within two
standard deviations are quoted as systematic uncertainties.

The uncertainty from knowledge of the LHCb detector's
length and momentum scale is negligible as is the
statistical uncertainty from the sWeights.
The tagging parameters are allowed to float in the fit using Gaussian constraints according to their uncertainties,
and thus their systematic uncertainties are propagated into the statistical uncertainties reported on
the physics parameters themselves.
The systematic uncertainties for $\phi_s$, $\Delta\Gamma_s$ and $\Gamma_s$ can be
treated as uncorrelated between this result and those in Ref.~\cite{LHCb-PAPER-2014-059}.

\section{Conclusions}
\label{sec:conclusions}

  Using a dataset corresponding to an integrated luminosity of $3.0
  \invfb$ collected by the LHCb experiment in $pp$ collisions during LHC
  Run 1, a flavour tagged, decay-time dependent angular analysis of approximately 4700
  $\Bs\to\psi(2S)\phi$ decays is performed.
  The analysis gives access to a number of physics parameters
 including the \CP-violating phase, average decay-width and decay-width difference
 of the \Bs system as well as the polarisation amplitudes and strong phases of the
 decay.
 The effective decay-time
  resolution and effective tagging power are approximately 47\fs and $3.9\%$, respectively.
 This is the first measurement of the \CP content of the  $\Bs\to\psi(2S)\phi$ decay
 and first time that $\phi_s$ and $\Delta\Gamma_s$ have been
 measured in a final state containing the $\psi(2S)$ resonance.
 The results are consistent with previous
measurements~\cite{Abazov:2011ry,CDF:2011af,LHCb-PAPER-2014-059,Aad:2016tdj,Khachatryan:2015nza},
the SM predictions~\cite{Charles:2004jd,Lenz:2006hd,Artuso:2015swg}, and show no evidence
of \CP violation in the interference between \Bs meson mixing and decay.
 The parameter $|\lambda|$ is consistent with unity,
 implying no evidence for direct \CP violation in $\Bs \to \psitwos \phi$ decays.
 The fraction of longitudinal polarisation in the $\Bs \to \psitwos \phi$ decay is measured to
 be lower than that in the $\Bs \to \jpsi \phi$ decay,
 consistent with the predictions of Ref.~\cite{Hiller:2013cza}.

\section*{Acknowledgements}

\noindent We express our gratitude to our colleagues in the CERN
accelerator departments for the excellent performance of the LHC. We
thank the technical and administrative staff at the LHCb
institutes. We acknowledge support from CERN and from the national
agencies: CAPES, CNPq, FAPERJ and FINEP (Brazil); NSFC (China);
CNRS/IN2P3 (France); BMBF, DFG and MPG (Germany); INFN (Italy);
FOM and NWO (The Netherlands); MNiSW and NCN (Poland); MEN/IFA (Romania);
MinES and FANO (Russia); MinECo (Spain); SNSF and SER (Switzerland);
NASU (Ukraine); STFC (United Kingdom); NSF (USA).
We acknowledge the computing resources that are provided by CERN, IN2P3 (France), KIT and DESY (Germany), INFN (Italy), SURF (The Netherlands), PIC (Spain), GridPP (United Kingdom), RRCKI and Yandex LLC (Russia), CSCS (Switzerland), IFIN-HH (Romania), CBPF (Brazil), PL-GRID (Poland) and OSC (USA). We are indebted to the communities behind the multiple open
source software packages on which we depend.
Individual groups or members have received support from AvH Foundation (Germany),
EPLANET, Marie Sk\l{}odowska-Curie Actions and ERC (European Union),
Conseil G\'{e}n\'{e}ral de Haute-Savoie, Labex ENIGMASS and OCEVU,
R\'{e}gion Auvergne (France), RFBR and Yandex LLC (Russia), GVA, XuntaGal and GENCAT (Spain), Herchel Smith Fund, The Royal Society, Royal Commission for the Exhibition of 1851 and the Leverhulme Trust (United Kingdom).

\addcontentsline{toc}{section}{References}
\ifx\mcitethebibliography\mciteundefinedmacro
\PackageError{LHCb.bst}{mciteplus.sty has not been loaded}
{This bibstyle requires the use of the mciteplus package.}\fi
\providecommand{\href}[2]{#2}

\newpage

\centerline{\large\bf LHCb collaboration}
\begin{flushleft}
\small
R.~Aaij$^{40}$,
B.~Adeva$^{39}$,
M.~Adinolfi$^{48}$,
Z.~Ajaltouni$^{5}$,
S.~Akar$^{6}$,
J.~Albrecht$^{10}$,
F.~Alessio$^{40}$,
M.~Alexander$^{53}$,
S.~Ali$^{43}$,
G.~Alkhazov$^{31}$
% ...

{\footnotesize \it
% ...
$ ^{5}$Clermont Universit{\'e}, Universit{\'e} Blaise Pascal, CNRS/IN2P3, LPC, Clermont-Ferrand, France\\
$ ^{6}$CPPM, Aix-Marseille Universit{\'e}, CNRS/IN2P3, Marseille, France\\
% ...
$ ^{10}$Fakult{\"a}t Physik, Technische Universit{\"a}t Dortmund, Dortmund, Germany\\
% ...
$ ^{31}$Petersburg Nuclear Physics Institute (PNPI), Gatchina, Russia\\
% ...
$ ^{39}$Universidad de Santiago de Compostela, Santiago de Compostela, Spain\\
$ ^{40}$European Organization for Nuclear Research (CERN), Geneva, Switzerland\\
% ...
$ ^{43}$Nikhef National Institute for Subatomic Physics, Amsterdam, The Netherlands\\
% ...
$ ^{48}$H.H. Wills Physics Laboratory, University of Bristol, Bristol, United Kingdom\\
% ...
$ ^{53}$School of Physics and Astronomy, University of Glasgow, Glasgow, United Kingdom\\
% ...
}
\end{flushleft}

\end{document}